\documentclass[aip,reprint, amsmath,amssymb]{revtex4-1}

\usepackage{graphicx}
\usepackage{dcolumn}
\usepackage{bm}

\begin{document}


\title{Knudsen layer formation in laser induced thermal desorption} 



\author{Akihiko Ikeda}\email[E-mail: ]{a-ikeda@iis.u-tokyo.ac.jp}
\author{Masuaki Matsumoto}
\author{Shohei Ogura}
\author{Tatsuo Okano}
\author{Katsuyuki Fukutani}\email[E-mail: ]{fukutani@iis.u-tokyo.ac.jp}
\affiliation{Institute of Industrial Science, The University of Tokyo, 4-6-1, Komaba, Meguro, Tokyo, 153-8505, Japan}


\date{\today}

\begin{abstract}
Laser induced thermal desorption of Xe atoms into vacuum from a metal surface following the nano-second pulsed laser heating was investigated by the time-of-flight (TOF) measurement. The desorption flow was studied at a wide range of desorption flux by varying the initially prepared Xe coverage $\Theta$ (1 ML $ = 4.5\times10^{18}$ atoms/m$^{2}$). At $\Theta = 0.3$ ML, the TOF of Xe was well represented by a Maxwell-Boltzmann velocity distribution, which is in good agreement with thermal desorption followed by collision-free flow. At $\Theta > 0.3$ ML, the peak positions of the TOF spectra were shifted towards the smaller values and became constant at large $\Theta$, which were well fitted with a shifted Maxwell-Boltzmann velocity distribution with a temperature $T_{\mathrm{D}}$ and a stream velocity $u$. With $T_{\mathrm{D}}$ fixed at 165 K, $u$ was found to increase from 80 to 125 m/ s with increasing $\Theta$ from 1.2 to 4 ML. At $\Theta > 4$ ML, the value of $u$ become constant at 125 m/ s. The converging feature of $u$ was found to be consistent with analytical predictions and simulated results based on the Knudsen layer formation theory. We found that the Knudsen layer formation in laser desorption is completed at Knudsen number Kn $<0.39$.

\end{abstract}

\pacs{47.45.-n, 47.70.Nd, 68.43.Nr, 68.43.Vx}

\maketitle 

\section{Introduction}
Laser induced desorption of atoms and molecules from solid surfaces is a vital phenomenon to investigate fundamentals such as surface electronic structures and dynamics.\cite{Ho1, Zheng, Wang2009, Klass, Toker} Besides, laser desorption is an essential technique for pulsed laser deposition used in thin film growth \cite{Kools1992, Kools1993, Konomi2009, Konomi2010, Singh, Kar, Peterlongo} and mass spectrometry of protein employing matrix-assisted laser desorption ionization.\cite{Puretzky} When desorption flux is small, the velocity distribution of desorbed atoms is directly governed by the desorption mechanism. When the desorption flux is large enough, on the other hand, the post-desorption collision between desorbed particles may become significant and modify the velocity distribution in the vicinity of the surface after the desorption.

Manifestations of the collision effect in laser induced desorption have been reported both by experiments and simulations as the modifications of the angular and velocity distribution of desorbing atoms and molecules.\cite{Cowin1978, NoorbatchaJCP1987, NoorbatchaPRB1987, NoorbatchaSS1988} Cowin \textit{et al.} investigated the angular dependence of the translational temperature of D$_{2}$ desorbed from tungsten surfaces under a pulsed laser irradiation.\cite{Cowin1978} The translational temperature of D$_{2}$ desorbed in the surface normal direction was higher than those in oblique directions. They attributed the variation of the translational temperature to the collision effect. Noorbatcha \textit{et al.} used the direct Monte Carlo simulation of desorbing atoms to investigate the collision effect. They showed that even in the sub-monolayer regime the collision noticeably modifies the final angular, velocity and rotational-energy distributions.\cite{NoorbatchaJCP1987, NoorbatchaPRB1987, NoorbatchaSS1988} However, there has not been any model that can quantitatively estimate the degree of modification by the post-desorption collision in laser desorption so far.
\begin{figure}[b]
\begin{center}
\includegraphics[scale=.7, clip]{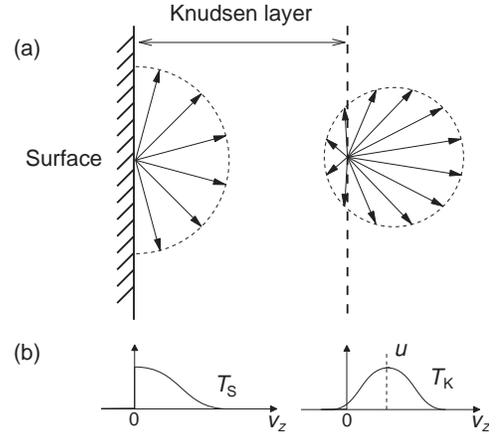} 
\caption{(a) A schematic sketch of the Knudsen layer formed above the surface following the laser induced thermal desorption. The desorption flux is assumed to be large enough for the post-desorption collisions to take place and to further form the Knudsen layer. The arrows denote the velocity vectors of desorbed atoms at the surface and at the end of the Knudsen layer. (b) Schematic plots of the velocity distribution of the desorbed atoms in the $z$ (surface normal) direction at the surface and at the end of the Knudsen layer. $T_{\mathrm{S}}$, $T_{\mathrm{K}}$ and $u$ denote the surface temperature, the translational temperature of the desorbed atoms at the end of the Knudsen layer and the stream velocity, respectively.\label{KL}}
\end{center}
\end{figure}

Knudsen layer formation theory has been developed in rarefied gas dynamics to model the steady flow of the strong evaporation from the surface.\cite{Ytrehus, Cercignani, Davidsson} As shown in Fig. \ref{KL}, the initial velocity distribution of thermally desorbed species at the surface is well described by a ''half-range'' Maxwell-Boltzmann velocity distribution.\cite{Ytrehus} In the Knudsen layer theory, as a result of intensive post-desorption collisions, the half-range velocity distribution at the surface is thermally equilibrated to a full-range Maxwell-Boltzmann velocity distribution with a stream velocity at some distance from the surface. This thermalization layer is defined as the Knudsen layer. The theory analytically predicts for monoatomic gas that the ratio of the translational temperature at the end of Knudsen layer $T_{\mathrm{K}}$ to the surface temperature $T_{\mathrm{S}}$ and the Mach number of the desorption flow at the end of the Knudsen layer become 0.65 and 1.0, respectively.\cite{KellySS1988, KellyNIMB1988}

Kelly and Dreyfus discussed that the Knudsen layer formation theory may be applicable to the pulsed desorption flow with a large desorption flux. However, it is not straightforward because the pulsed desorption involves complex time evolution of the density and velocity distributions of desorbed atoms.\cite{Stein} Sibold and Urbassek have shown by means of the Monte Carlo simulation of the Boltzmann equation that the pulsed desorption flow at an intense flux is well characterized by the above values predicted by the Knudsen layer formation theory. Although previous experimental studies recognized the collision effect in laser desorption,\cite{Taborek, Burgess, Cowin1985, HusslaBBPC1986, HusslaCJP1986} the formation of the Knudsen layer in laser desorption has been discussed only by theory and simulations.\cite{KellySS1988, KellyNIMB1988, Sibold1991} So far, any experimental confirmation of Knudsen layer formation in laser desorption has not been presented. For an experimental verification of the theory, a systematic observation of the translational temperature and stream velocity of the desorption flow as a function of the desorption flux is strongly required.

In the present paper, we investigated the laser induced thermal desorption (LITD) of Xe from an Au(001) surface by means of the time-of-flight (TOF) measurement as a function of the wide range of desorption flux by varying the initial Xe coverage $\Theta$. We found that at $\Theta$ close to 0 ML, the TOF was well analyzed by a Maxwell-Boltzmann velocity distribution. Hence, the desorption at $\Theta$ close to 0 ML is rationalized by the thermal desorption followed by the collision-free flow. At $\Theta$ close to monolayer, we observed that the peak positions of the TOF spectra shift towards smaller values. Assuring that the desorption is only thermally activated, we regard this modification of the TOF as the manifestation of the collision effect. At larger $\Theta$, the peak positions of the TOF spectra become constant. We deduced the Mach number $M$ of the desorption flow to be 0.96 at large $\Theta$ under the assumption that $T_{\mathrm{K}}/T_{\mathrm{S}}=0.65$. The obtained value of $M$ and the saturating behavior of u at $\Theta > 4$ ML both well coincide with the previously reported values by the Knudsen layer theory and simulations. The facts suggest the formation of the Knudsen layer in LITD at large $\Theta$. Furthermore, we tentatively estimated the Knudsen number of the
initial desorption flow.


\begin{figure}
\begin{center}
\includegraphics[scale=0.7, clip]{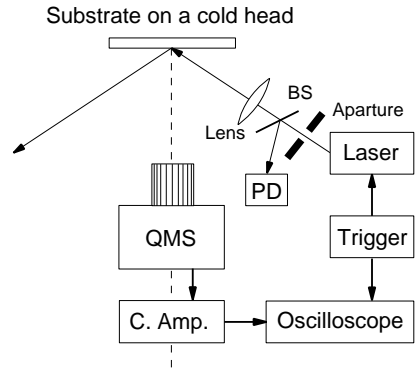} 
\caption{A schematic drawing of the experimental setup for the time-of-flight measurement of Xe following the laser induced thermal desorption from Au surfaces. C. Amp., PD and BS denote fast current amplifier, photo diode and beam splitter, respectively.\label{exp}}
\end{center}
\end{figure}

\section{Experiment}
The Au specimen and the physisorbed Xe layers were prepared in the following manner. An Au disk with 001 orientation was cut out from a single-crystal rod and was chemically and mechanically polished. The Au disk was put into an ultra high vacuum (UHV) chamber with a base pressure of $2\times 10^{-8}$ Pa. The sample surface was cleaned by several cycles of Ar$^{+}$ sputtering and annealing at 700 K in the UHV by the electron bombardment. The cleanliness of the specimen was confirmed by the observation of a clear reconstructed (5$\times$20) pattern \cite{VanHove} by low-energy electron diffraction and no contamination in the Auger electron spectrum. The sample temperature was monitored with a Chromel-Alumel type thermocouple directly spot-welded to the side of the Au disk. The disk was cooled down and kept at 23 K by a closed-cycle He compressor type refrigerator during all LITD experiments.

The Xe gas was dosed onto the Au surface by backfilling the chamber. The Xe coverage $\Theta$ was monitored by the LITD yield. The LITD yield showed a linear dependence on the Xe exposure in the present experimental condition. The saturation coverage at the sample temperature of 80 K was defined to be 1 ML $=4.5\times10^{18}$ atoms/m$^{2}$ as in Ref. \onlinecite{Mcelhiney}. $\Theta$ was varied by varying the dosage of Xe gas onto the Au surface. We note that the desorption flux is defined as $\Theta$ devided by  the mean desorption period $\tau$. On the basis of the result employing the LITD simulation, $\tau$ was shown to be constant at about 4 ns at $\Theta$ concerned in the present study.\cite{Ikeda} Hence, the desorption flux can be varied by varying $\Theta$.

\begin{figure}
\begin{center}
\includegraphics[scale=0.5, clip]{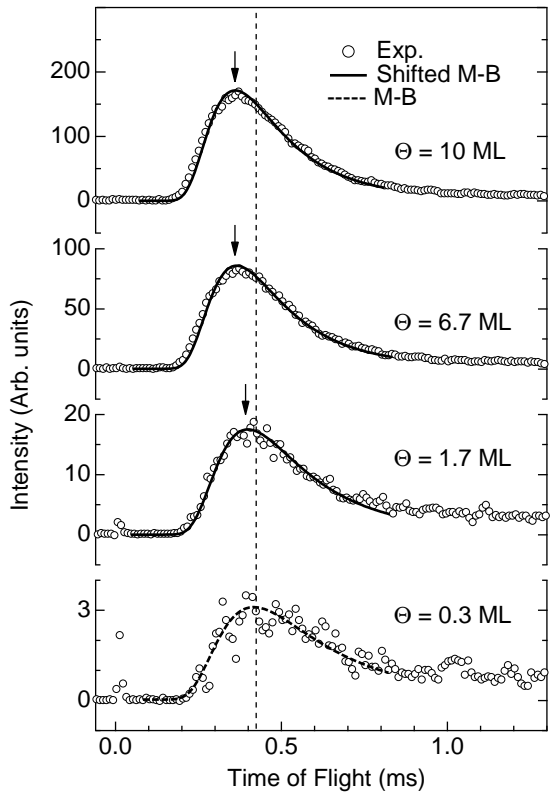} 
\caption{Time-of-flight (TOF) spectra of Xe from Au surfaces following pulsed laser irradiations. The Xe coverage $\Theta$ are 10, 6.7, 1.7 and 0.3 ML from top to bottom, respectively. The dashed curves and the solid curves are the Maxwell-Boltzmann (M-B) velocity distribution and the shifted M-B velocity distribution fitted to the experimental results, respectively. The vertical dashed line and the arrows indicate the peak position of each spectrum. \label{tof}}
\end{center}
\end{figure}

The TOF measurement of the LITD of Xe from an Au surface was carried out in the following manner. The experimental setup is schematically shown in Fig. \ref{exp}. An ArF excimer laser (Lambda Physik) was used as a pulsed laser source. The wavelength was 193 nm and the time duration was 8 ns. The incidence angle was 25$^{\circ}$ from the surface normal direction. The irradiated area on the sample disk was $\sim$0.1 cm$^{2}$. The laser pulse was guided on to the surface through an aperture and a quartz window. The laser power absorbed by the sample was fixed at 80 mJ/cm$^{2}$ in all experiments taking the reflectivity into consideration. At this laser power absorbed by the sample, the LITD of Xe is the dominant desorption process as shown in Ref. \onlinecite{Ikeda}. The reflected light escapes the UHV chamber through another quartz window. A quadrupole mass spectrometer (QMS: Balzers QMA125) was placed in the surface normal direction at a distance of 10 cm from the Au specimen. The QMS was tuned for a high-sensitive detection of Xe with a low mass selectivity. The ion current of the QMS was amplified with a fast current amplifier (Keithley 427) and sent to an oscilloscope (Tektronixs: TDS620B). The desorption of Xe from the Au(001) surface was induced with only one laser pulse in each LITD experiment. The respective TOF was recorded using the oscilloscope with a single acquisition mode.

\begin{figure}[!t]
\begin{center}
\includegraphics[scale=0.5, clip]{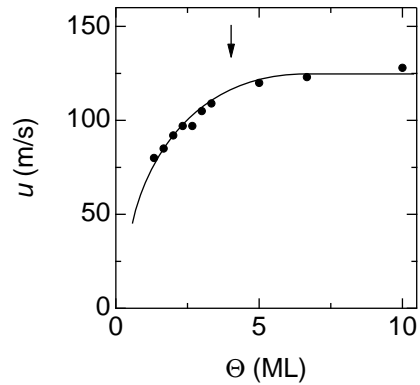}
\caption{The stream velocity $u$ of the desorbed Xe atoms from Au surfaces following the pulsed laser irradiations as a function of the Xe coverage $\Theta$. The data are obtained by analyzing the time-of-flight spectrum with the shifted Maxwell-Boltzmann velocity distribution. The solid line is a guide for eyes. $u$ become constant at above 4 ML which is denoted by the arrow.\label{tud}}
\end{center}
\end{figure}

\section{Results}
We obtained a series of TOF spectra of desorbed Xe from Au surfaces following pulsed laser irradiation as shown in Fig. \ref{tof}. The TOF spectrum of Xe from the Au surface at $\Theta = 0.3$ ML is shown at the bottom of Fig. \ref{tof}. The vertical dashed line at 0.42 ms shows the peak position of the spectrum. The spectrum was well analyzed with a Maxwell-Boltzmann velocity distribution in a flux from a thermal source described as, \cite{Comsa}
\begin{equation}
J(v)dv=Av^{3}\exp\left(-\frac{mv^{2}}{2kT_{\mathrm{D}}}\right)dv,\label{mb}
\end{equation}
where $m$, $k$ and $v$ are the mass of a Xe atom, the Boltzmann constant and the velocity of Xe atoms, respectively, and $A$ and $T_{\mathrm{D}}$ are fitting parameters. In the analysis, we convert Eq. (\ref{mb}) to a TOF function for the density sensitive detector, which results in the form $f(t)dt=a t^{-4}\exp(-bt^{-2})dt$. By fitting Eq. (\ref{mb}) to the TOF of Xe at $\Theta=0.3$ ML, we obtained $T_{\mathrm{D}}$ of 255 K as illustrated by the dashed curve at the bottom of Fig. \ref{tof}.

At the top and middle of Fig. \ref{tof}, the TOF spectra of desorbed Xe at $\Theta=10$ to 1.7 ML are shown. We note that the peak positions of those TOF spectra, which are indicated by the arrows, are shifted towards smaller values with increasing $\Theta$ compared with that of $\Theta$ = 0.3 ML. It was found that the peak positions of the TOF spectra become constant at about 0.34 ms at $\Theta\geq5.0$ ML.

\section{Discussion}

\subsection{Desorption mechanism}

We first discuss that the observed desorption proceeds only via thermal activation. In the previous study,\cite{Ikeda} we investigated the same system by varying the photon fluence at $\Theta=1$ ML. At the small fluence region where the laser power is below 20 mJ/cm$^{2}$, we observed a desorption of Xe only at a photon energy of 6.4 eV and no desorption at 2.3 eV. We assigned this photodesorption as a non-thermal desorption via a transient Xe$^{-}$ formation. At a large fluence region where laser power exceeds 60 mJ/cm$^{2}$ we observed, on the other hand, that thermal desorption is the dominant desorption process regardless of the photon energy. These observations are confirmed by comparing the experimentally observed $T_{\mathrm{D}}$ and Xe desorption yield with the results of LITD simulations. Therefore, we regard that the initial desorption is only thermally activated in the present experiment.

\subsection{Analysis of TOF spectra}

In order to analyze the TOF spectra at $\Theta>0.3$ ML, we introduce a shifted Maxwell-Boltzmann velocity distribution which is characterized by $u$. Generally, the velocity distribution in a desorption flux shows an angular distribution and is expressed by the elliptical distribution which involves an angular dependent translational temperature as described in Ref. \onlinecite{SiboldJAP1993}. In the present experiment, the detector was fixed in the surface normal direction. Therefore, the translational temperature only in this direction is detected in the present experiment. In this case, the shifted Maxwell-Boltzmann velocity distribution in a flux from a thermal source is expressed as,\cite{KellySS1988, Sibold1991}
\begin{equation}
J(v)dv=Av^{3}\exp\left\{-\frac{m(v-u)^{2}}{2kT_{\mathrm{D}}}\right\}dv.\label{smb}
\end{equation}
Under the condition of $u=0$, Eq. (\ref{smb}) is identical to Eq. (\ref{mb}). In the analysis, Eq. (\ref{smb}) was also converted to a TOF function for the density sensitive detector. 

The TOF spectra at $\Theta > 0.3$ ML were well analyzed using Eq. (\ref{smb}). However, we found it difficult to obtain a unique set of $T_{\mathrm{D}}$ and $u$. The spectra were well reproduced with an arbitrary value of $T_{\mathrm{D}}$ from 165 to 310 K by adjusting $u$ from 125 to 0 m/s. We note that, with a fixed value of $T_{\mathrm{D}}$ between 165 and 310 K, the obtained value of $u$ monotonically increases with increasing $\Theta$ from 0.3 to 4 ML and becomes constant at $\Theta > 4$ ML. For example, Fig. \ref{tud} shows u as a function of $\Theta$ obtained by using Eq. (\ref{smb}) with $T_{\mathrm{D}}$ fixed at 165 K, of which the fitting curves are shown as solid curves in Fig. \ref{tof}.

\subsection{Moderate desorption at $\Theta\simeq0$ ML}

At $\Theta\simeq0$, the observed feature of the TOF can be rationalized with a model employing thermal desorption followed by collision-free flow. At the instance of thermal desorption, the desorbed Xe gas is in thermal equilibrium with the surface at the temperature $T_{\mathrm{S}}$. Here, it is reasonable to consider the half-range Maxwell-Boltzmann velocity distribution for desorbed Xe expressed as \cite{KellySS1988}
\begin{align}
f(\mathbf{v})d\mathbf{v}=A\exp\left(-\frac{m\mathbf{v}^{2}}{2kT_{\mathrm{S}}}\right)d\mathbf{v},&\notag\\
v_{z}>0& \label{mbv}
\end{align}
where $A$ and $\mathbf{v}$ are a normalization factor and the velocity vector of Xe atoms in the Cartesian coordinate, respectively. We note that in Eq. (\ref{mbv}) the velocity component in the $z$ direction has a distribution only at $v_{z}>0$,\cite{KellySS1988} assuming the $z$ axis to be the surface normal direction. Given the desorption period is much shorter than the flight time and the irradiation diameter is much smaller than the flight distance, Eq. (\ref{mbv}) can be, as a velocity distribution in a flux from a thermal source, transformed to the same form as Eq. (\ref{mb}). Thus, we notice that $T_{\mathrm{D}}=T_{\mathrm{S}}$ at $\Theta\simeq0$.\cite{Wedler, Cowin1978}

For a qualitative analysis, we estimated the surface temperature during the present LITD experiment using the first order desorption rate equation, assuming that the desorption activation energy is 240 meV\cite{Mcelhiney} and that the desorption period is 4 ns.\cite{Wedler, Brand} The calculated results show that the required surface temperature is 260 K, which is in good agreement with the obtained value of $T_{\mathrm{S}}=255$ K at $\Theta=0.3$ ML. The agreement indicates that the model of the thermal desorption followed by the collision-free flow well describes the observed TOF at $\Theta\simeq0$, in good accordance with the results by Cowin \cite{Cowin1978} and Wedler \cite{Wedler}.

\subsection{Intensive desorption at $\Theta>4$ ML}

At $\Theta>4$ ML, we discuss that the desorption flow is described as an intense flow, where the post-desorption collision modifies the velocity distribution.\cite{NoorbatchaPRB1987, NoorbatchaJCP1987} Several theoretical studies have speculated that the so-called Knudsen layer is formed in the vicinity of the surface as a result of collisions.\cite{KellySS1988, KellyNIMB1988, KellyPRA1992, Sibold1991} As schematically drawn in Fig. \ref{KL}, in the Knudsen layer model, the initial velocity distribution is in thermal equilibrium with the surface at $T_{\mathrm{S}}$ except that there is no distribution at $v_{z}<0$ as in Eq. (\ref{mbv}). As a result of a significant number of post-desorption collisions, the half-range velocity distribution at $z=0$ becomes thermally equilibrated at some distance from the surface to a full-range Maxwell-Boltzmann velocity distribution with a stream velocity $u$.\cite{Ytrehus, Cercignani, KellySS1988} The model also requires that, at the end of the Knudsen layer, the temperature of the flow becomes identical ($T_{\mathrm{K}}$) in all degrees of freedom. Hence, at the end of the Knudsen layer, the velocity distribution may be described as \cite{Ytrehus, Cercignani, KellySS1988}
\begin{equation}
f(\mathbf{v}) d\mathbf{v}=A\exp\left[-\frac{m\{v_{x}^{2}+v_{y}^{2}+(v_{z}-u_{\mathrm{K}})^{2}\}}{2kT_{\mathrm{K}}}\right] d\mathbf{v},\label{smbv}
\end{equation}
where $v_{i}$, $u_{\mathrm{K}}$ and $T_{\mathrm{K}}$ are the velocity component in the Cartesian coordinate, stream velocity and the temperature at the end of the Knudsen layer, respectively. Under the condition that the flight distance is sufficiently longer than the thickness of the Knudsen layer besides the conditions described above, Eq. (\ref{smbv}) is converted, as a velocity distribution from a thermal source, to the identical form to Eq. (\ref{smb}). Thus, we see that $T_{\mathrm{D}}=T_{\mathrm{K}}$ and $u=u_{\mathrm{K}}$ at $\Theta>4$ ML. Equation (\ref{smb}) was shown to well describe the experimental results of the TOFs at $\Theta>4$ ML. This fact, along with the convergence feature of $u$, indicates that the experimental result at $\Theta>4$ ML may be rationalized with the Knudsen layer formation model.

\subsection{Mach number of the flow}

We further examine the results at $\Theta>4$ ML by quantitatively comparing the obtained results of $u_{\mathrm{K}}$ with previous theoretical reports on Knudsen layer formation in the steady strong evaporation \cite{Ytrehus, Cercignani} and the pulsed desorption \cite{Sibold1991} from plane surfaces. We compare $u_{\mathrm{K}}$ in terms of the Mach number $M$ at the end of the Knudsen layer. $M$ is defined as
\begin{equation}
M=\frac{u_{\mathrm{K}}}{c}=u_{\mathrm{K}}\sqrt{\frac{m}{\gamma kT_{\mathrm{K}}}},\label{mach}
\end{equation}
where $c$ and $\gamma$ are the local velocity of sound and the heat capacity ratio of the gas, respectively, the latter of which is 5/3 for monoatomic gas as in the present case. 
\begin{table}
\caption{\label{tablem}Mach number $M$ obtained in the present study and previously reported values by simulation and theory on the Knudsen layer.
}%
\begin{ruledtabular}
\begin{tabular}{lccc}
Method&
$M$&
Condition\\
\colrule
Experiment\footnote{Present study}& 0.96 & Pulsed flow\\
Simulation \cite{Sibold1991} & 1.0 & Pulsed flow\\
Simulation \cite{SiboldPFFD1993} & 1.0 & Steady flow\\
Theory \cite{Ytrehus, Cercignani, KellySS1988} & 0.99 & Steady flow\\
\end{tabular}
\end{ruledtabular}
\end{table}

Ytrehus \cite{Ytrehus} and Cercignani \cite{Cercignani} formulated $M$ and $T_{\mathrm{K}}/T_{\mathrm{S}}$ by finding a solution to the Boltzmann equation under the Knudsen layer formation model assuming the conservation of the particle number, momentum and the energy flux between the surface and the end of the Knudsen layer. Sibold and Urbassek estimated $M$ and $T_{\mathrm{K}}/T_{\mathrm{S}}$ using the Monte Carlo simulation of the Boltzmann equation in one-dimension for both pulsed flow \cite{Sibold1991} and steady flow \cite{SiboldPFFD1993} conditions. Those reports have stated that at the end of the Knudsen layer, $M$ should always be close to unity and that $T_{\mathrm{K}}/T_{\mathrm{S}}$ is at around 0.65.

As discussed in IV. B, it was difficult to unambiguously determine 
the value of $T_{\mathrm{D}}$. Here, we assume the relation of $T_{\mathrm{K}}/T_{\mathrm{S}}=0.65$ following the theoretical studies and evaluate the Mach number. With the value of $T_{\mathrm{S}} = 255 $ K, we obtain the value of $T_{\mathrm{K}}$ to be 165 K. With $T_{\mathrm{K}} = 165 $ K and Eq. (\ref{mach}), we obtain the value of $M$ to be 0.96. As can be seen in Table I, the previously reported values of $M$ show a quantitative agreement with the obtained value of $M$ in the present study. This indicates that the observed trend of $u$ at $\Theta > 4$ ML is consistent with the Knudsen layer formation theory.\cite{KellySS1988}

With adiabatic expansion, $M$ should well exceed unity.\cite{Kelly1990, KellyPRA1992, KellyNIMB1992} We note that we observed the saturation of $M$ at around unity at $\Theta>$ 4 ML in the present experimental conditions. This indicates that the adiabatic expansion is practically absent, although the slight increase of $u$ at $\Theta>$ 4 ML may be due to the adiabatic expansion.

\subsection{Knudsen number of the flow}
In order to generalize the present result, we tentatively consider mean gas density $\bar{n}$, mean free path $\lambda$ and Knudsen number Kn in the vicinity of the surface at the moment of desorption as a function of $\Theta$. Kn is defined by \cite{Sibold1991}
\begin{equation}
\mathrm{Kn}=\frac{\lambda}{l_{z}}=\frac{1}{\sqrt{2}\bar{n}\sigma l_{z}},\label{KN}
\end{equation}
where $l_{z}$ and $\sigma$ are the representative length of the system and the Van der Waals collision cross section of Xe ($1.0\times10^{-19}$ m$^{2}$), respectively.

Here, we simply regard that $l_{z}$ is the mean thickness of the gas cloud above the surface. It, then, reads that $l_{z}=\bar{v}_{z}\tau$, where $\bar{v}_{z}$ is the mean thermal velocity of desorbed Xe in the $z$ direction and $\tau$ is the mean desorption period. We further take $\bar{v}_{z}=\sqrt{2kT_{\mathrm{S}}/\pi m}$ and estimate that $\bar{n}=\Theta/l_{z}$. By substituting $\bar{n}$ in Eq. (\ref{KN}), we obtain a simple relation $\mathrm{Kn}=a/(\sqrt{2}\tilde{\Theta})$ as introduced in Ref. \onlinecite{Sibold1991}. $\tilde{\Theta}$ and $a$ are the relative coverage $\tilde{\Theta}=\Theta/\Theta_{\mathrm{S}}$ with the monolayer saturation coverage $\Theta_{\mathrm{S}}$ and the ratio of the area occupied by an atom at $\Theta_{\mathrm{S}}$ to $\sigma$ described as $a=1/(\sigma\Theta_{\mathrm{S}})$, respectively.

Now, we see that Kn depends simply on $\tilde{\Theta}$ and $a$. Therefore, we can discuss the required condition for the formation of Knudsen layer in LITD for general systems in terms of Kn. We fixed $\tau=4$ ns. On the basis of the result employing the LITD simulation, $\tau$ showed little dependence on $\Theta$ in the region concerned in the present study.

In Table \ref{tablek}, we summarized the obtained $\bar{n}$, $\lambda$ and Kn as a function of $\Theta$. Kn can also be understood as an inverse of mean collision times per each atoms. In the present study, we observed the formation of the Knudsen layer at $\Theta>4$ ML, which corresponds to Kn $<$ 0.39 and to more than $\sim2.6$ collisions per desorbing atom. The manifestation of collision effects is observed at $\Theta>0.3$ ML, which corresponds to Kn $>5.2$. The results coincide with the previous theoretical work by Sibold and Urbassek \cite{Sibold1991} that the Knudsen layer is formed at $\Theta=2.5$ ML and that the collision effect already appears while the Knudsen layer is not formed at $\Theta=0.25$ ML, respectively. The obtained values of Kn also agree well with the previous experimental observation of the collision effect in LITD of D$_{2}$/W at $\Theta=1$ ML reported by Cowin \textit{et al.} \cite{Cowin1978}. They also coincide with the theoretically estimated collision number of 2.9 per atom at $\Theta=1$ ML reported by Noorbatcha \textit{et al.}.\cite{NoorbatchaJCP1987, NoorbatchaPRB1987}

\begin{table}

\caption{\label{tablek}Mean gas density $\bar{n}$, mean free path $\lambda$ and Knudsen number Kn in the vicinity of the surface at the moment of laser desorption as a function of $\Theta$.
}%
\begin{ruledtabular}
\begin{tabular}{cccc}
$\Theta$ (ML)&
$\bar{n}$ (10$^{24}$ atoms/m$^{3}$)&
$\lambda$ (nm)&
Kn\\
\colrule
0.3&3.3&2114&5.2\\
4.0&45&159&0.39\\
10&111&63&0.16\\
\end{tabular}
\end{ruledtabular}
\end{table}

Lastly, we note the observed feature at $\Theta\simeq1$ ML, where the Knudsen layer is not formed. Thus, the velocity distributions are not necessarily well fitted with Eq. (\ref{smb}). Nevertheless, they were very well analyzed with Eq. (\ref{smb}) in all $\Theta$ at $\Theta\simeq1$ ML. The result is in good agreement with the simulated results by Sibold and Urbassek,\cite{Sibold1991} although any analytical formulation for the flow at $\Theta\simeq1$ ML has not been presented so far. Hence, further theoretical and experimental works are required for elucidating the mechanism for the modification of velocity distribution at $\Theta\simeq1$ ML due to the post-desorption collision as pointed out in Ref. \onlinecite{Sibold1991}.

\section{Conclusion}
In conclusion, we observed the TOFs of Xe from a Au(001) surface by LITD at a wide range of $\Theta$. At $\Theta\simeq0.3$ ML, the TOF is well analyzed by the Maxwell-Boltzmann velocity distribution at 255 K indicating thermal desorption followed by the collision-free flow. At $\Theta>0.3$ ML, the TOF is well analyzed by the shifted Maxwell-Boltzmann velocity distribution, which is regarded as the manifestations of the post-desorption collision effects. At $\Theta>4$ ML, the value of $u$ under the assumption that $T_{\mathrm{K}}/T_{\mathrm{S}}=0.65$ became constant at around 125 m/s, which corresponds to $M=0.96$, being in good agreement with the prediction by the Knudsen layer formation theory. In the LITD experiment, we found that the collision effect already appears at Kn $\simeq5.2$ and that the Knudsen layer is formed at Kn $<0.39$, which corresponds to a mean collision number of greater than 2.6 per atom for general systems.

\begin{acknowledgments}
This work was supported by Grant-in Aid for Scientific Research (A) of Japan Society for the Promotion of Science (JSPS). A. I. acknowledges support from a Research Assistant of Global COE Program "The Physical Sciences Frontier", MEXT, Japan and TEPCO Memorial Foundation.
\end{acknowledgments}

\end{document}